\documentclass[useAMS,usenatbib]{mn2e-1}
\usepackage{amsfonts}
\usepackage{amsmath}
\usepackage{amssymb,url}
\usepackage{myaasmacros}
\usepackage{pbox}

\def\bea{\begin{eqnarray}}
\def\ena{\end{eqnarray}}

\usepackage[dvips]{graphicx}
\usepackage{amsmath,amssymb}
\usepackage{myaasmacros}
\usepackage{color}
\usepackage{multirow}

\title[Revisiting variable gamma-ray sky at 1 GeV ...]{Revisiting variable gamma-ray sky at 1 GeV with 6 years of
  Fermi-LAT data}
\author[M.S. Pshirkov \& G.I. Rubtsov]{M.S. Pshirkov$^{1,2,3}$\thanks{E-mail: pshirkov@sai.msu.ru}, G.I. Rubtsov$^{2}$\thanks{E-mail: grisha@ms2.inr.ac.ru }\\
$^{1}$Sternberg Astronomical Institute, Lomonosov Moscow State University, Universitetsky prospekt 13, 119992, Moscow, Russia\\
$^{2}$Institute for Nuclear Research of the Russian Academy of Sciences, 117312, Moscow, Russia\\
$^{3}$Pushchino Radio Astronomy Observatory, 142290 Pushchino, Russia\\
}

\begin{document}

\date{}


\maketitle

\label{firstpage}

\begin{abstract}
We perform a  blind search for the variability of the
$\gamma$-ray sky in the energy range $E > 1$ GeV using 308 weeks of
the Fermi-LAT data. We use the technique based on the comparison of
the weekly photon counts and exposures in sky pixels by means of the
Kolmogorov-Smirnov test. We consider the flux variations in the region
significant if statistical probability of uniformity is less than
$4\times10^{-6}$, which corresponds to 0.05 false detections in the
whole set of 12288 pixels. Close inspection of the detected variable regions result
in identification of 8 sources without previous known variability. Two
of them are included in the second Fermi LAT source catalogue (FBQS J122424.1+243623 and GB6 J0043+3426) and one (3EG J1424+3734) 
was reported by EGRET and also was included in the First Fermi LAT source catalogue
(1FGL), but is missing in  the 2FGL. Possible identifications of  five
other sources are obtained using NED
and SIMBAD databases (1RXS J161939.9+765515, PMN J2320-6447, PKS 0226-559, PKS J0030-0211, PMN J0225-2603). These new variable gamma-ray sources demonstrate recurring
flaring activity with time scale $\sim$ weeks and have hard spectra.  Their spectral energy distributions deviate significantly from a simple power-law shape and often peak around $\sim$GeV. These properties of activity are typical for flaring blazars.

\end{abstract}




\section{Introduction}
\label{sec:intro}

In the previous paper \citep{Pshirkov2013} (Paper I) we developed a
method for the blind search of long time-scale variability in the
Fermi-LAT data.  We binned the gamma-ray sky into a large number of
relatively small pixels ($\sim3.6$ sq.deg.) and, using Kolmogorov-Smirnov test,   compared a cumulative distribution functions of the registered photons and  exposure. If there is no significant variability in a pixel these distributions should be close.  The statistically significant deviation of
the flux from uniform distribution is considered as an indication of
the presence of some variable source in the corresponding pixel. The
method is sensitive to the slight gradual changes in the flux that are
more difficult to discover with the alternative techniques. However,
that comes at a price of somehow reduced sensitivity to short,
burst-like events.

In the Paper I 168 weeks of data were analyzed, 151 pixels with a
significant variability were found (see Section \ref{sec:data} for the
definition of significance). Variability in these pixels was due to
 118 sources (some sources contributed to several neighboring
pixels), 27 out of which had not been previously reported as variable
 in the Astronomer's
Telegrams\footnote{http://www.astronomerstelegram.org/}(ATels), on the
Fermisky blog\footnote{http://fermisky.blogspot.com/} or in the
dedicated papers.

In this paper we extend the data set by 140 weeks to the first 308
weeks of the Fermi mission livetime. This reanalysis is justified not
only because of  almost  twofold increase in  the statistics but also by using the new
reconstruction framework Pass 7 Reprocessed (P7REP) recently presented
by the Fermi-LAT collaboration \citep{P7REP}. The new improved
instrument response functions (IRFs) may result in a change of
calculated exposures and thus significantly affect the results of our
analysis.

\section{Data and method}
\label{sec:data}

In this work we use the LAT Pass 7 Reprocessed weekly all-sky data
publicly available at the Fermi mission
website\footnote{http://fermi.gsfc.nasa.gov/ssc/data/access/}. The
analysis spans the time period of 308 weeks from August 04, 2008 to
June 23, 2014, corresponding to the mission elapsed time (MET) from
239557417~s to 425179643~s. We use the 'SOURCE' event class photons
with $E>1$~GeV and impose an Earth relative zenith angle cut of
$100^\circ$ and rocking angle cut of $52^\circ$.  The data selection
criteria and our method are identical to the ones described in details
in Paper I, here we only briefly repeat the main points:
\begin{itemize}
 \item  We bin  weekly data files using \textsc{HEALPIX} package
\citep{Healpix}  into a map of resolution $N_{\mathrm{side}} = 32$ in
galactic coordinates with 'RING' pixel ordering.  Total number of
pixels is equal to 12288 and the area of each pixel is 3.6 sq. deg,
chosen according to the size of Fermi-LAT point-spread function (PSF)
above 1 GeV  which is approximately $1^{\circ}$.
\item We estimate an integral weekly exposure for each pixel using the
  standard Fermi-LAT tools {\it gtltcube} and {\it gtexpcube} from
  the Science Tools v9r32p5 package.

 For each pixel we count the number of photons in each of 308 weeks
 and consider corresponding values of weekly exposure.  Cumulative
 distribution functions (CDFs) $\mathcal{P}(t),\mathcal{E}(t)$ for
 both photon counts and exposure are constructed.
\item We compare distributions in each pixel using the
  Kolmogorov-Smirnov test and find all pixels with a probability
  $P_{\mathrm{KS}}$ lower than the threshold value
  $P_0=4\times10^{-6}$. This threshold leads to an average 0.05 false
  detections in the blind search through the whole 12288 pixel set.
\end{itemize}

We have produced a dedicated set of Monte-Carlo simulated photons
with the {\it gtobssim} utility using the positions and spectra  of sources from the 2FGL catalogue\citep{2FGL} and the latest templates for the  diffuse
Galactic and isotropic gamma-ray radiation. All sources were naturally simulated as non-variable. Repeating the analysis on events from
this Monte-Carlo set we arrive at the lowest $P_{\mathrm{KS}}$ value
around $10^{-4}$  in total agreement  with the expectations.

\section{Results}
\label{sec:results}
The analysis resulted in 231 pixels with $P_{\mathrm{KS}}<P_0$. We
compare our findings with the results of the Paper I and concentrate
mainly on the differences.  First, there are 20 pixels reported in the
Paper I, but not detected now. These changes could be due to two
main causes: 1) variability may decay on longer timescales and thus
the corresponding probability for this timescale would increase
and, eventually, would overshoot the chosen threshold 2) refined exposure calculation with the  improved IRFs may affect an apparent pixel variability obtained with the older IRFs.
 The most intriguing case was the one of the pixel 5637 which
contains the Geminga pulsar. We have reported
$P_{\mathrm{KS}}=2.3\times10^{-7}$ for it in the Paper I. However in our new analysis this
probability was only $9\times10^{-5}$. This decrease is mostly attributed to the change of the
reconstruction -- the probability for the first 178 weeks recalculated
using P7REP gives only $P_{\mathrm{KS}}=4.5\times10^{-4}$.

Second, we deal with the pixels that have not been reported in the
Paper I (100 pixels total). In order to crudely locate a source within the
pixel to check whether its variability was already known,  we plot the photons belonging to the  weeks with the highest
flux and find the center of  their clustering.  After that, the attribution
was first made using 2FGL catalogue and then using
SIMBAD\footnote{http://simbad.u-strasbg.fr/simbad/} or
NED\footnote{http://ned.ipac.caltech.edu/} databases if no
counterpart was found in the 2FGL catalogue. 

Finally we focus on the pixels where no sources with previously known
variability was found. That comprises both pixels with known gamma-ray sources that were supposed to be non-variable and pixels with some previously unknown gamma-ray sources not included in the 2FGL catalogue. Therefore we discard the pixels containing  sources with variability
reported in ATels, on the Fermisky blog, or in the papers e.g.
\citep{FAVA2013} and \citep{Abdo2010}, or, finally, if their 
variability indices in the 2FGL catalogue exceed the 'official' threshold
value 41.6. In the end it left us with 8 pixels out of 100 corresponding to 8
different sources.

\begin{figure}
\begin{center}
(a)\includegraphics[width=5.5cm, angle=0]{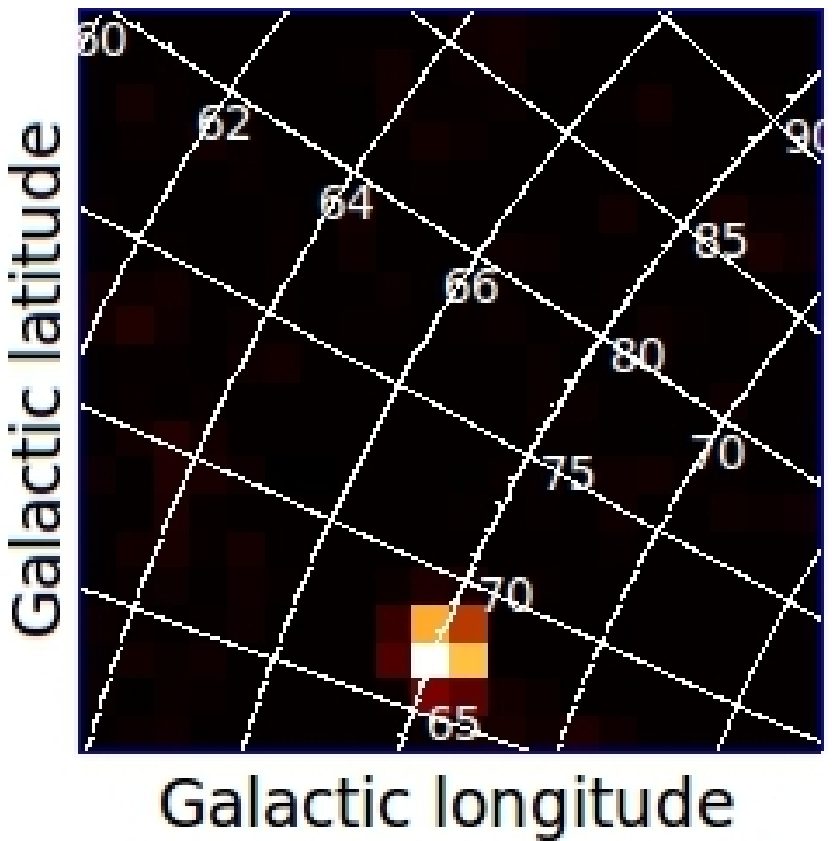} 
\includegraphics[width=.25cm, angle=0]{1pix.eps}
(b)\includegraphics[width=5.5cm, angle=0]{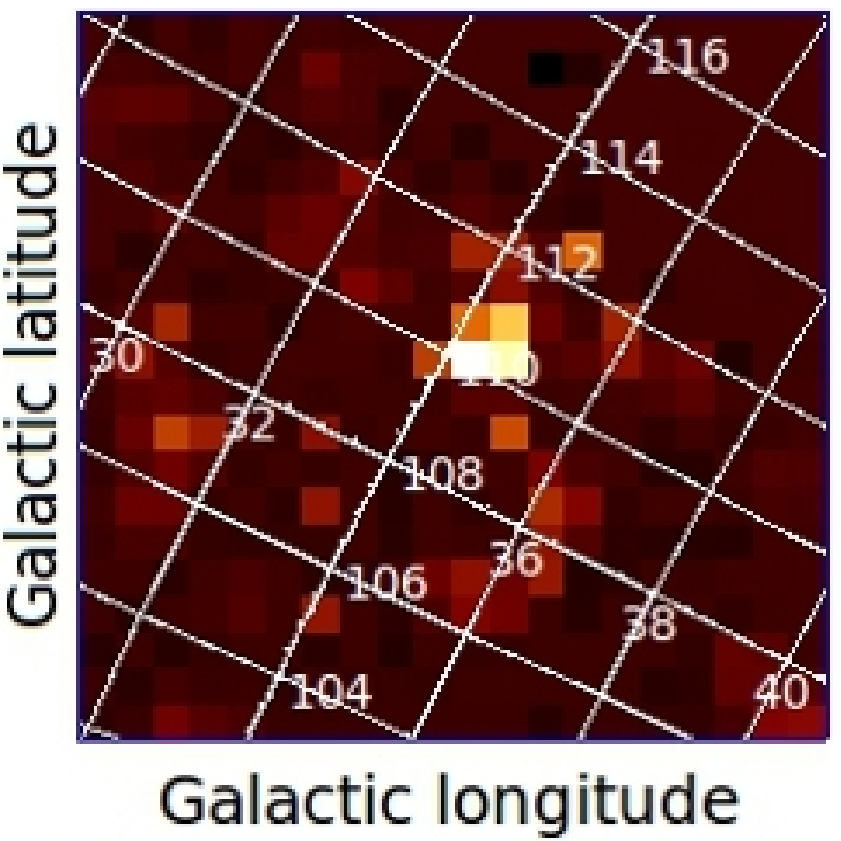} \\
\includegraphics[width=.25cm, angle=0]{1pix.eps}\\
(c)\includegraphics[width=5.5cm, angle=0]{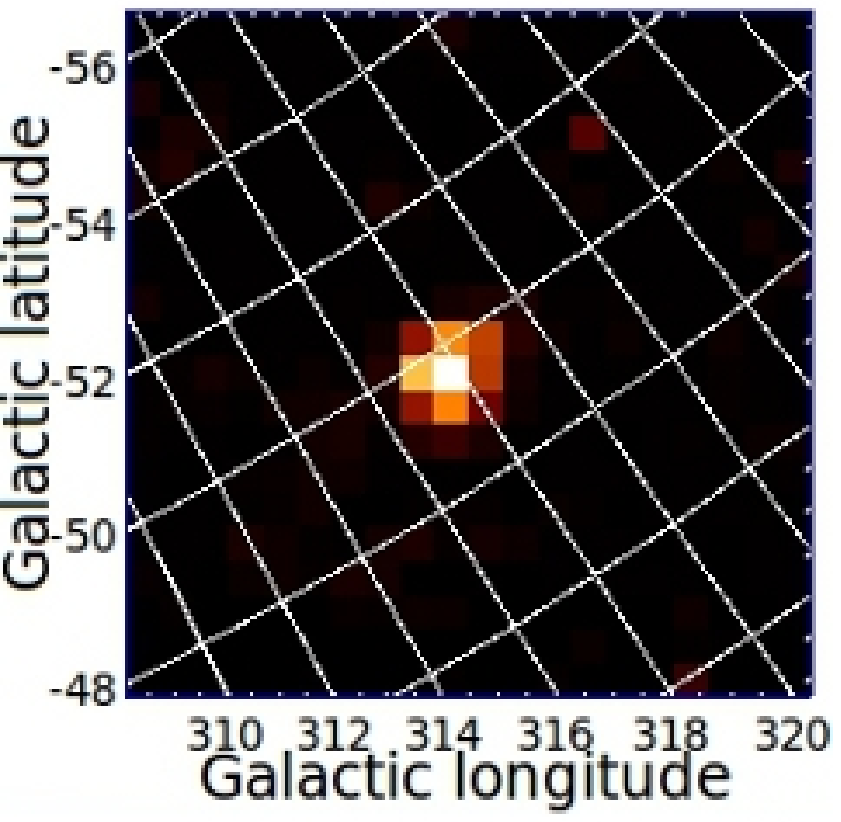} 
\includegraphics[width=.25cm, angle=0]{1pix.eps}
(d)\includegraphics[width=5.5cm, angle=0]{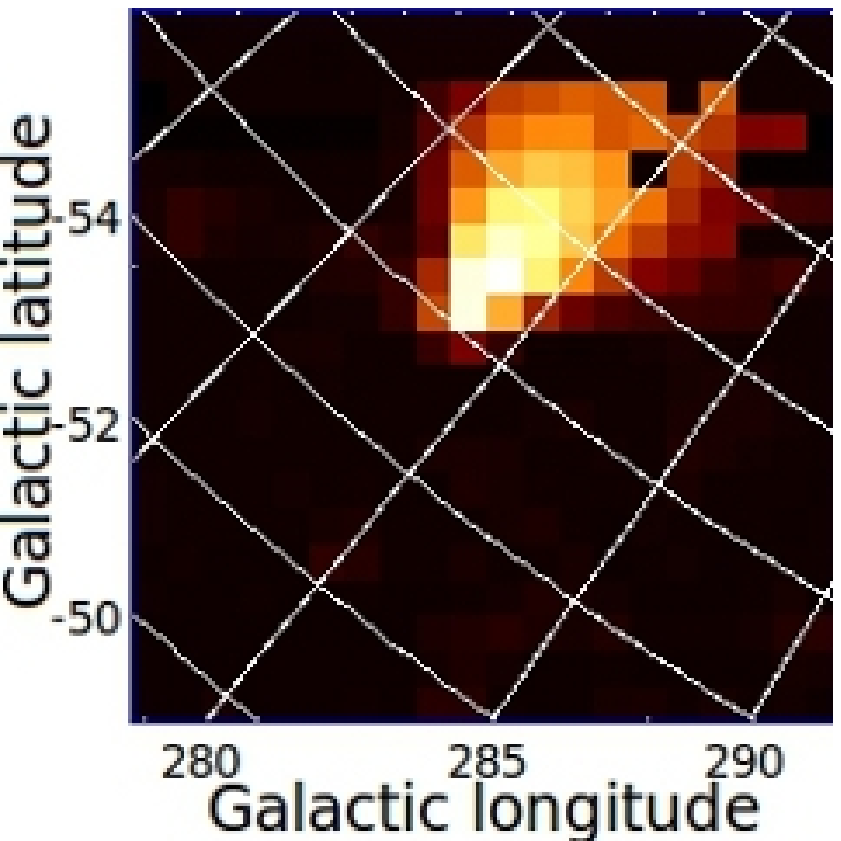} \\
\includegraphics[width=.25cm, angle=0]{1pix.eps}\\
(e)\includegraphics[width=5.5cm, angle=0]{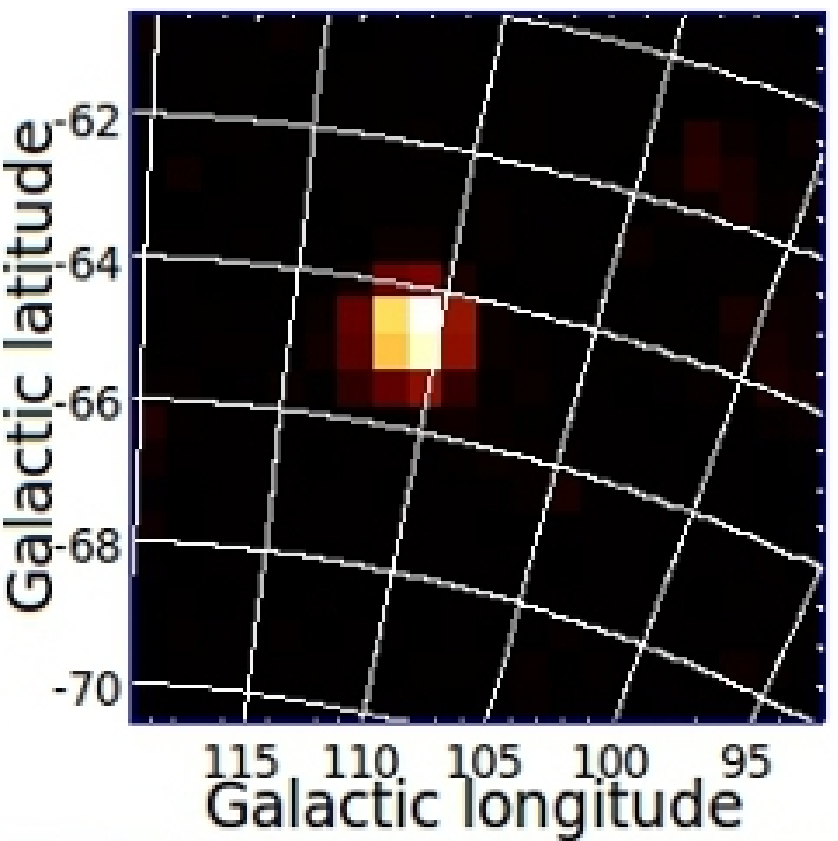} 
\includegraphics[width=.25cm, angle=0]{1pix.eps}
(f)\includegraphics[width=5.5cm, angle=0]{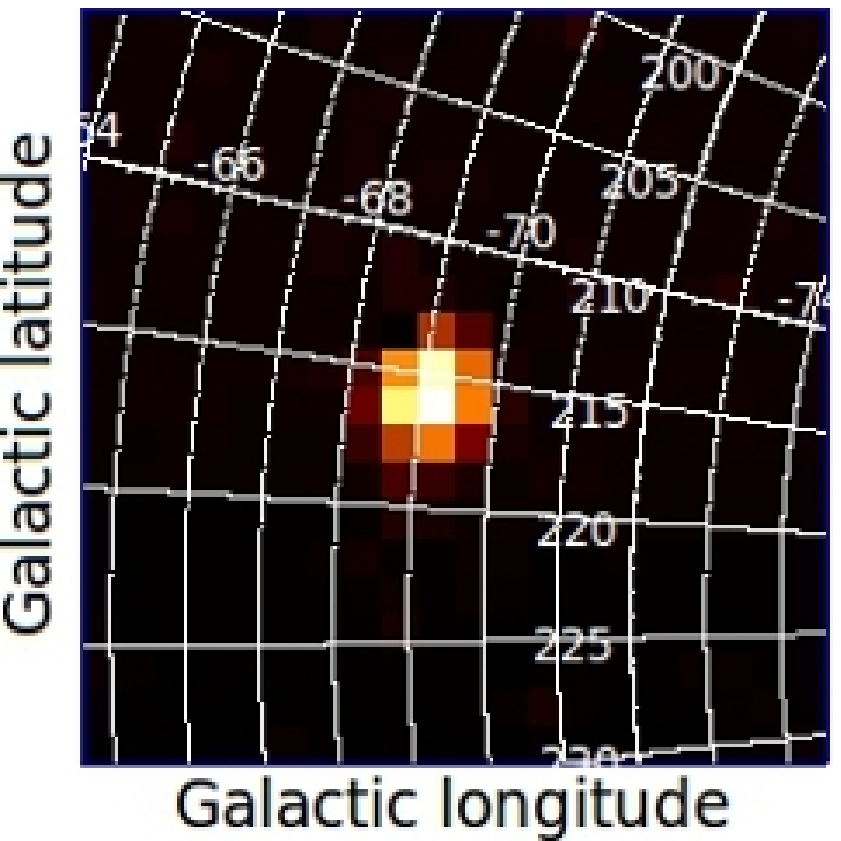} \\
\end{center}
\caption{TSMaps for the pixels listed in the Table \ref{table:undetected}. $(a)$ Pixel 431: $TS_{max}$=115 (weeks 66,293,301,302,316); $(b)$ Pixel 2663: $TS_{max}$=17 (weeks 109, 245, 250, 260, 261, 306); $(c)$  Pixel 10871: $TS_{max}$=64 (weeks 142, 170, 311, 314, 315); $(d)$ Pixel 11255: $TS_{max}$=126 (weeks 139, 238,276, 283, 285); $(e)$ Pixel 11696: $TS_{max}$=112 (weeks 200,201,276); $(f)$  Pixel 11901: $TS_{max}$=102 (weeks 134,135,138,184,289) } \label{fig:tsmap}
\end{figure}

With 8 remaining  pixels we perform more thorough search in the
following manner: first, we find a location of the apparent excess
using the \textit{gttsmap} utility. Known 2FGL sources and
the Galactic and isotropic diffuse components were included in the
initial sky model  that was provided as an input for \textit{gttsmap}. In 
two cases (pix.34, pix. 9067) no new significant excess was found
meaning that the flares occurred in the sources from the 2FGL
catalogue. These sources were identified by their large TS values in the
results of standard \textit{gtlike} fit. Variability in
these pixels was caused by 2FGL sources FBQS J122424.1+243623 (MS 1221.8+2452) and
GB6 J0043+3426 correspondingly. The first one is BL Lac type object at the redshift $z=0.218$
that was recently observed also in the TeV energy range by the MAGIC
collaboration \citep{MS1221}, the latter one is a blazar at the
redshift $z=0.966$.
 In the remaining six cases
(see Fig. \ref{fig:tsmap}) we have used
\textit{gtfindsrc} in order to refine the positions of the sources
(see Table \ref{table:positions}).

We have performed searches in the NED and SIMBAD databases using the best fit
positions and reported errors obtained earlier. One of the sources may
be identified with a high confidence: the reported variability in the
431-th pixel was caused by the 3EG J1424+3734 BL Lac type object that
was included into the 3rd EGRET catalogue \citep{3EGRET} and also into the first Fermi LAT catalogue (1FGL)
\citep{1FGL}. However it is absent in the 2FGL catalogue and that
clearly indicates its long-time variability. The other five sources were linked with the known radio- and X-ray sources from the NED and SIMBAD databases. One of the five, PKS 0226-559 (pix.11255), is a blazar with a  redshift $z=2.464$ \citep{CGRABS}. Two other sources, PKS J0030-0211 (pix. 11696) and  PMN J0225-2603 (pix. 11901\footnote {There is a possibility that the variability in this pixel was caused by another source, 1RXS J022517.5-260926 located only 6.6 arcmin away  from our best localization of the source but  a flat-spectrum radio source  PMN J0225-2603 looks a  more plausible candidate.}) are flat-spectrum radio sources. Two remaining candidate sources  1RXS J161939.9+765515 (pix. 2663) and PMN J2320-6447 (pix. 10871) has no classification type. 
	
\begin{figure}
\begin{center}

(a)\includegraphics[width=6.0 cm]{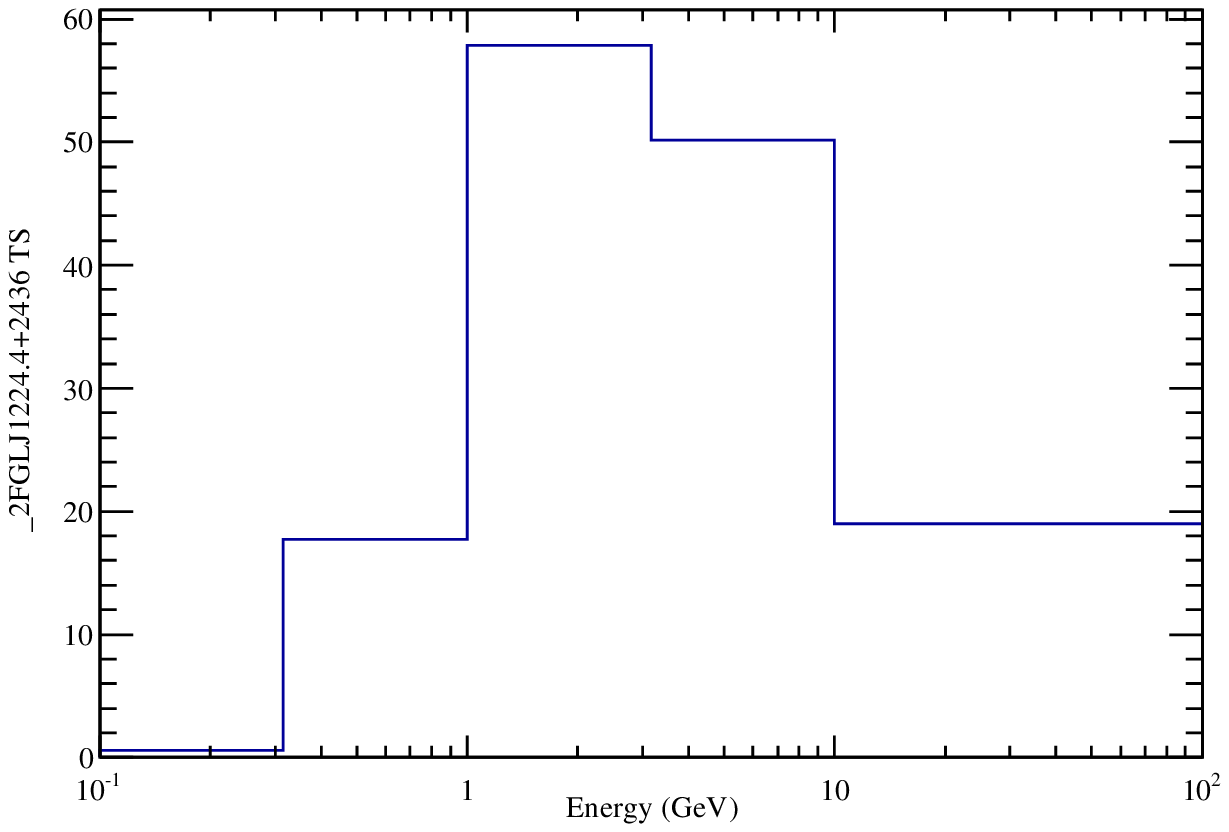} \includegraphics[width=6.0 cm]{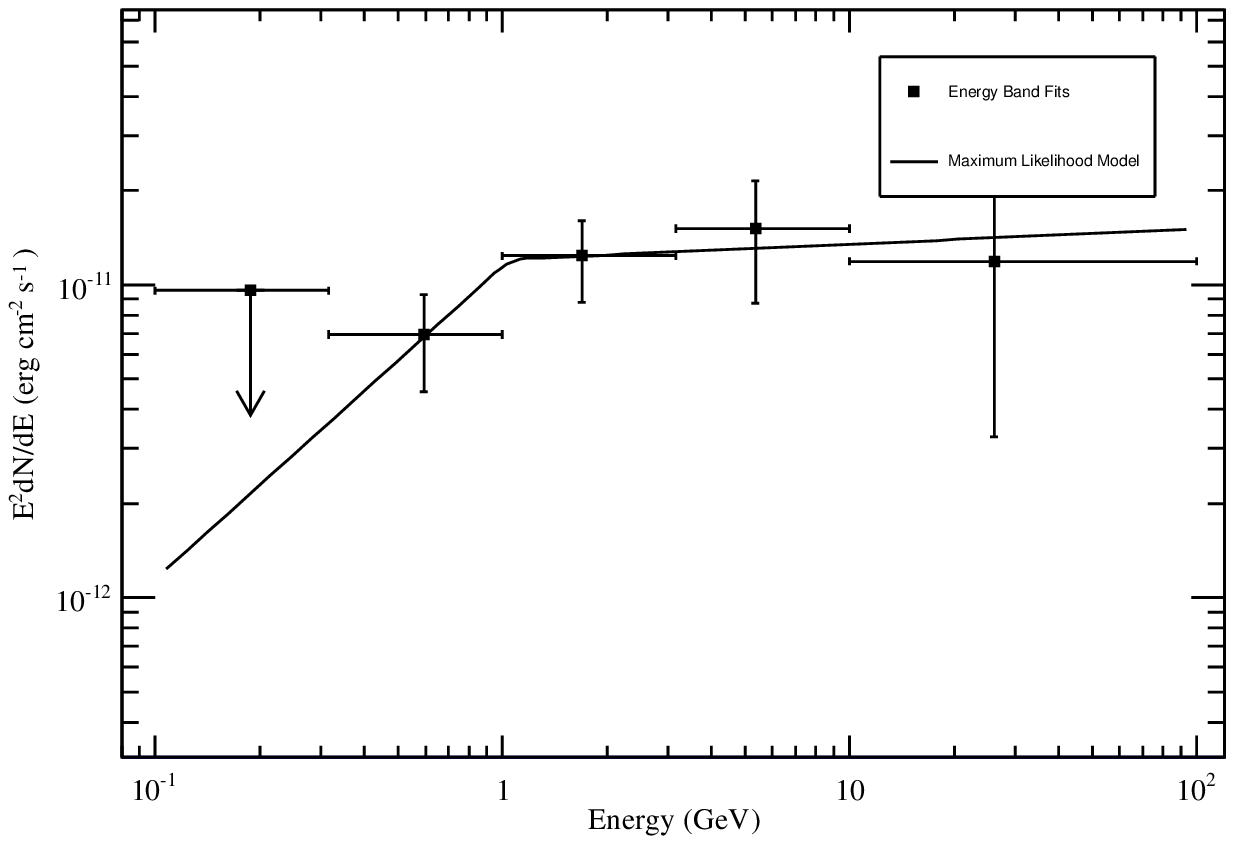}\\
(b)\includegraphics[width=6.0 cm]{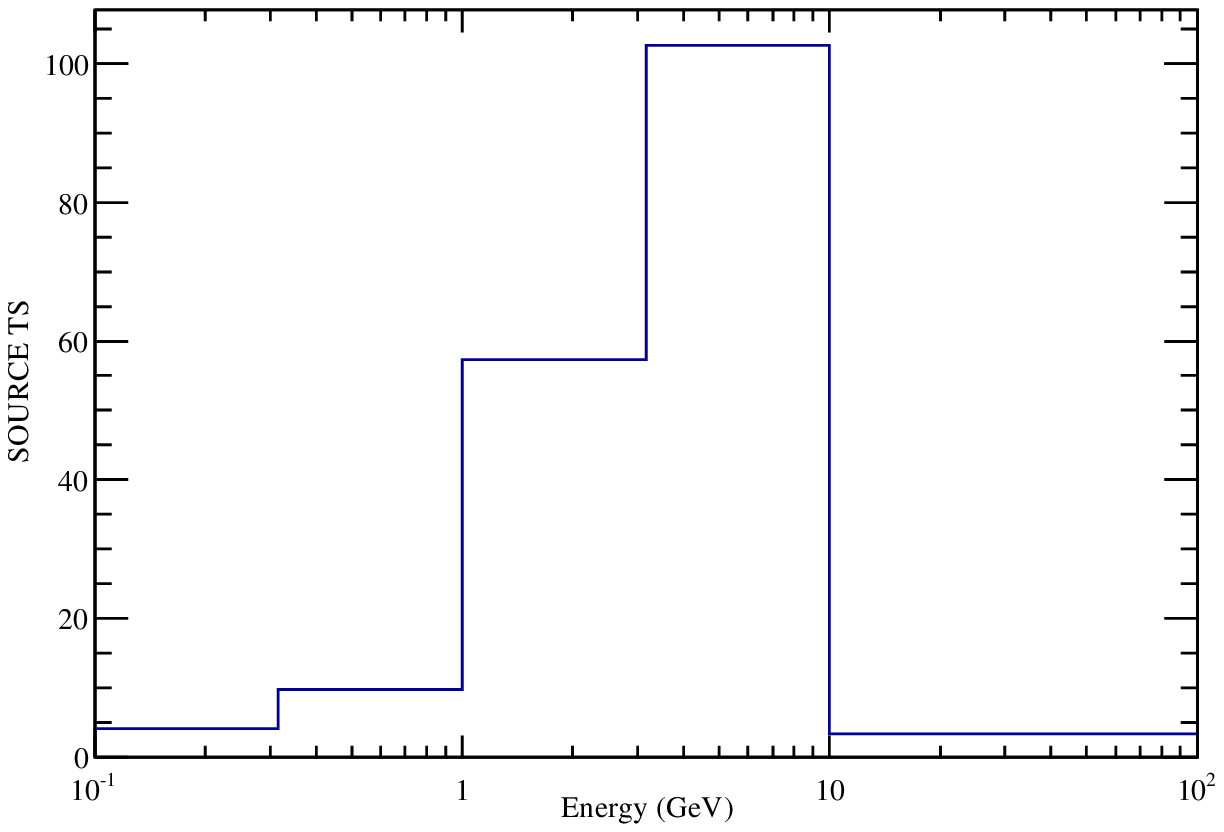} \includegraphics[width=6.0 cm]{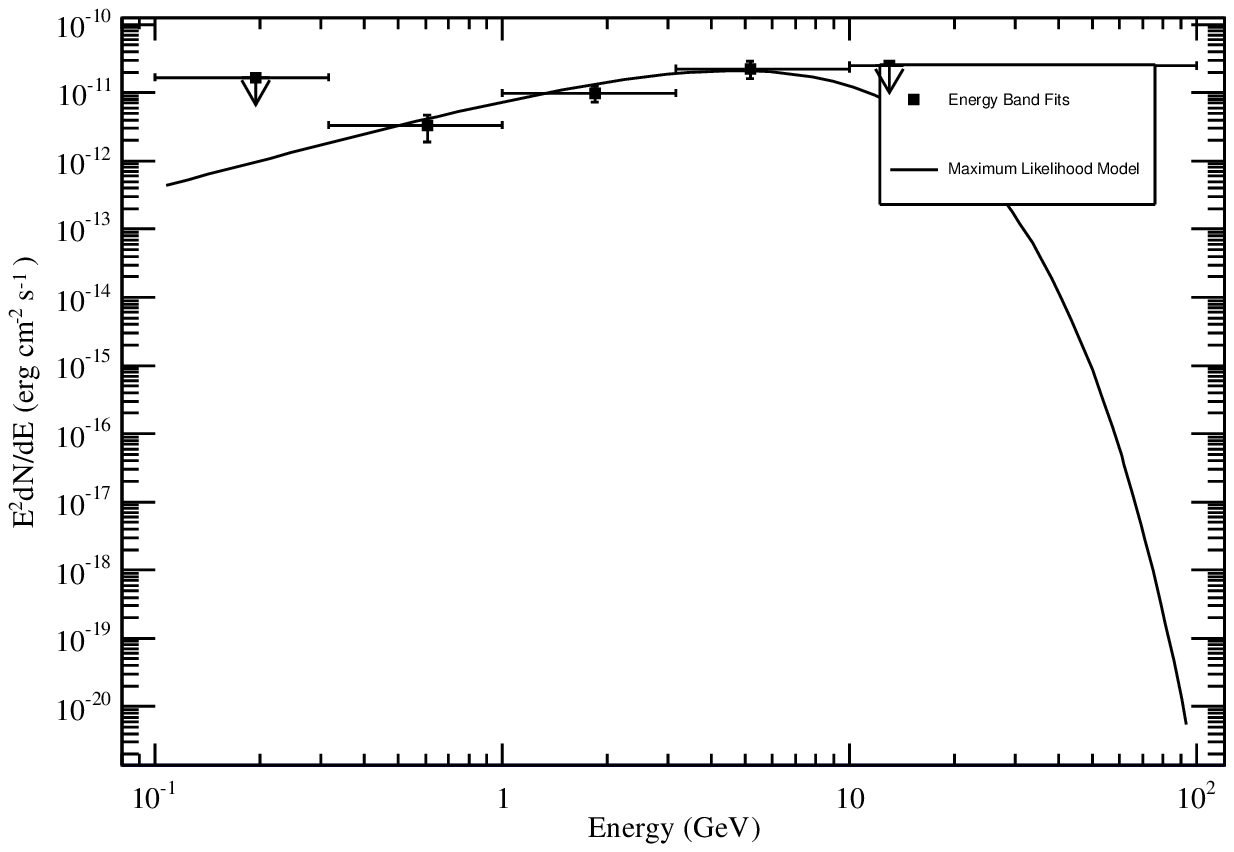}\\
(c)\includegraphics[width=6.0 cm]{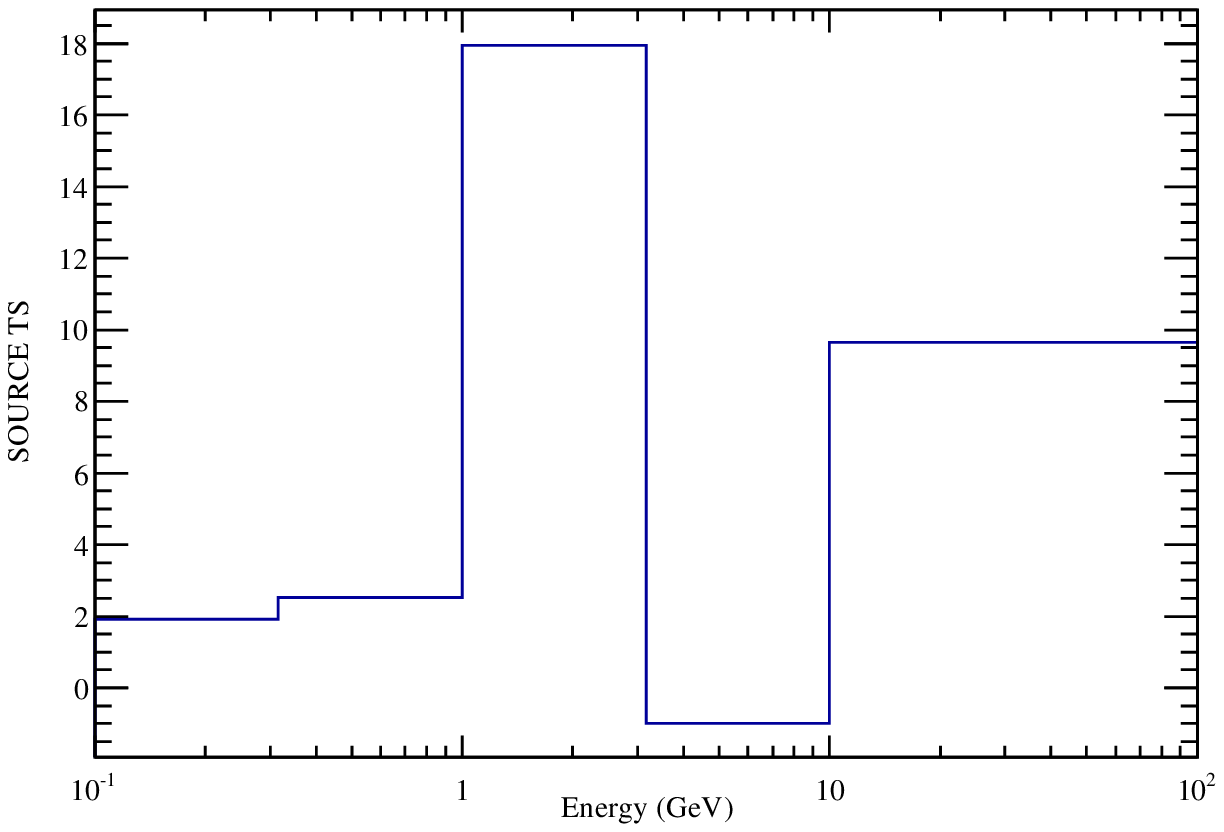} \includegraphics[width=6.0 cm]{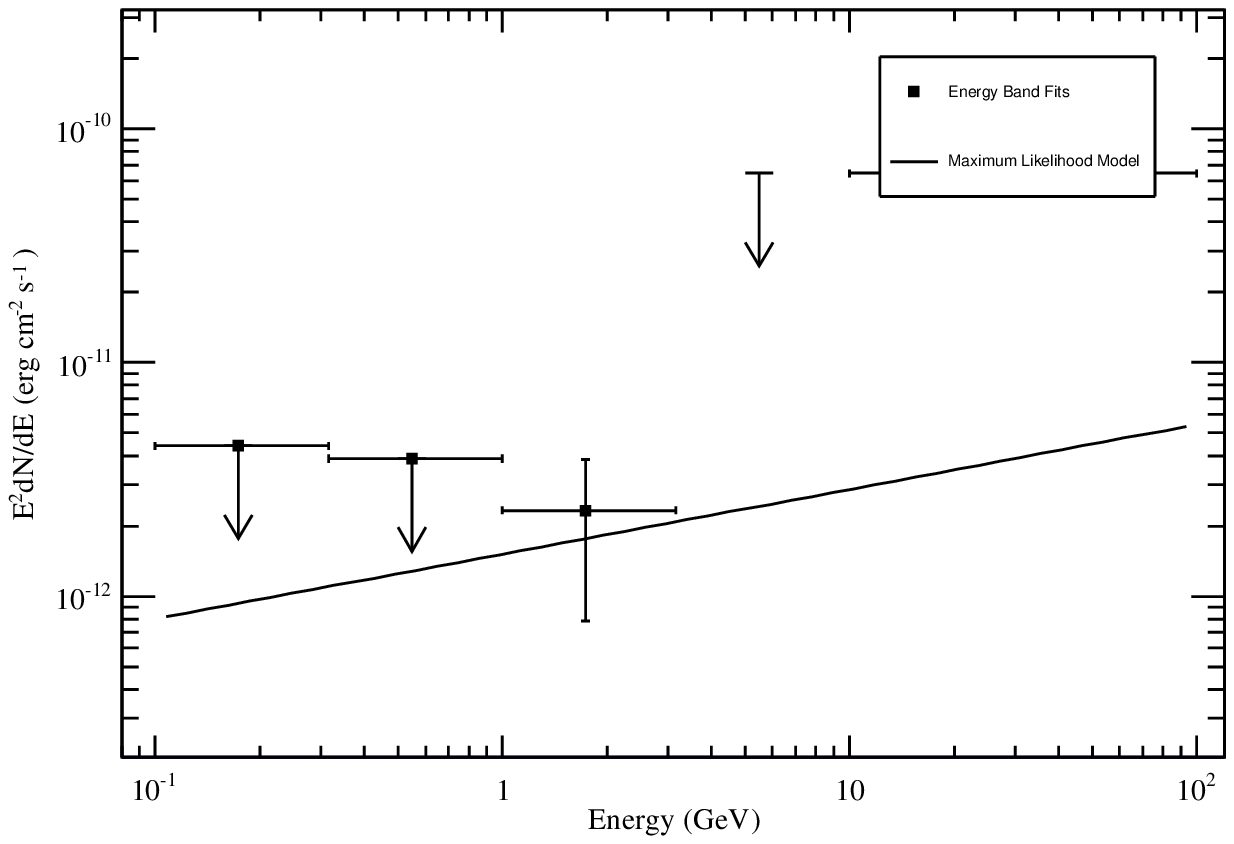}\\
(d)\includegraphics[width=6.0 cm]{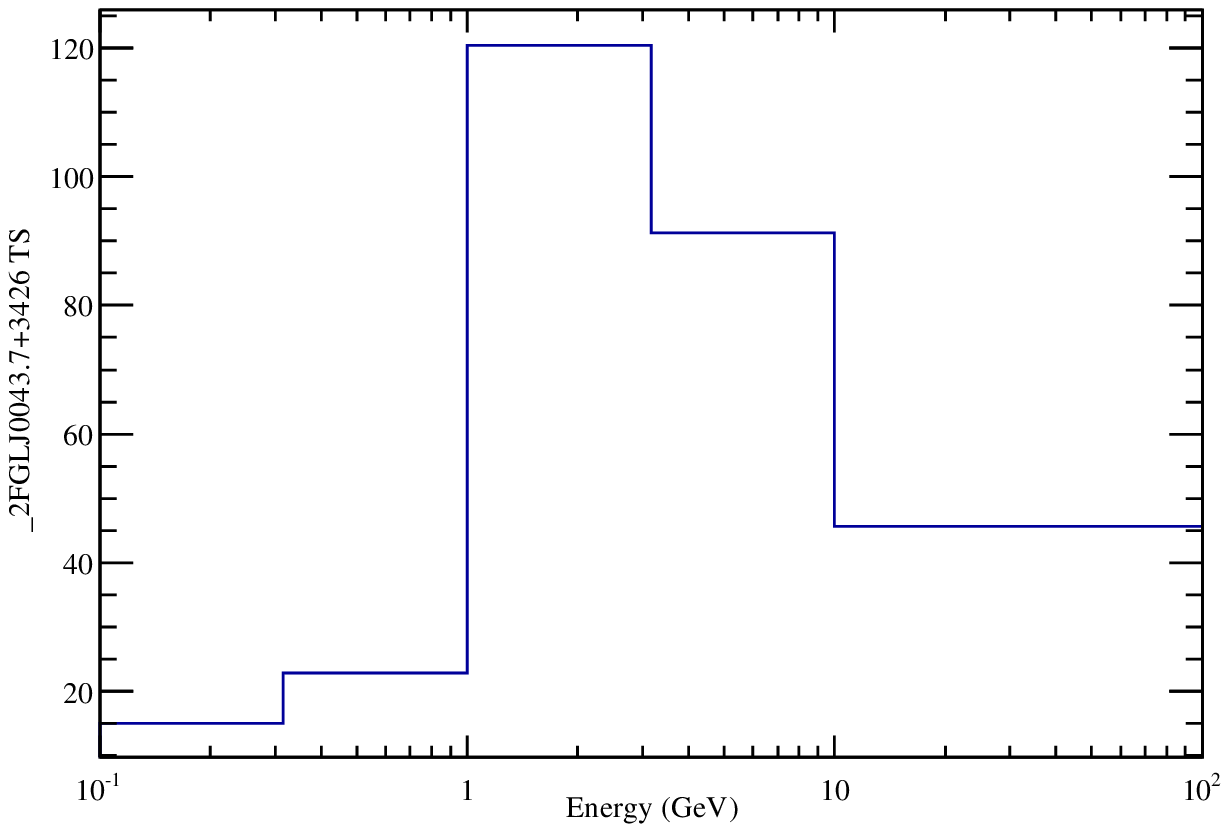} \includegraphics[width=6.0 cm]{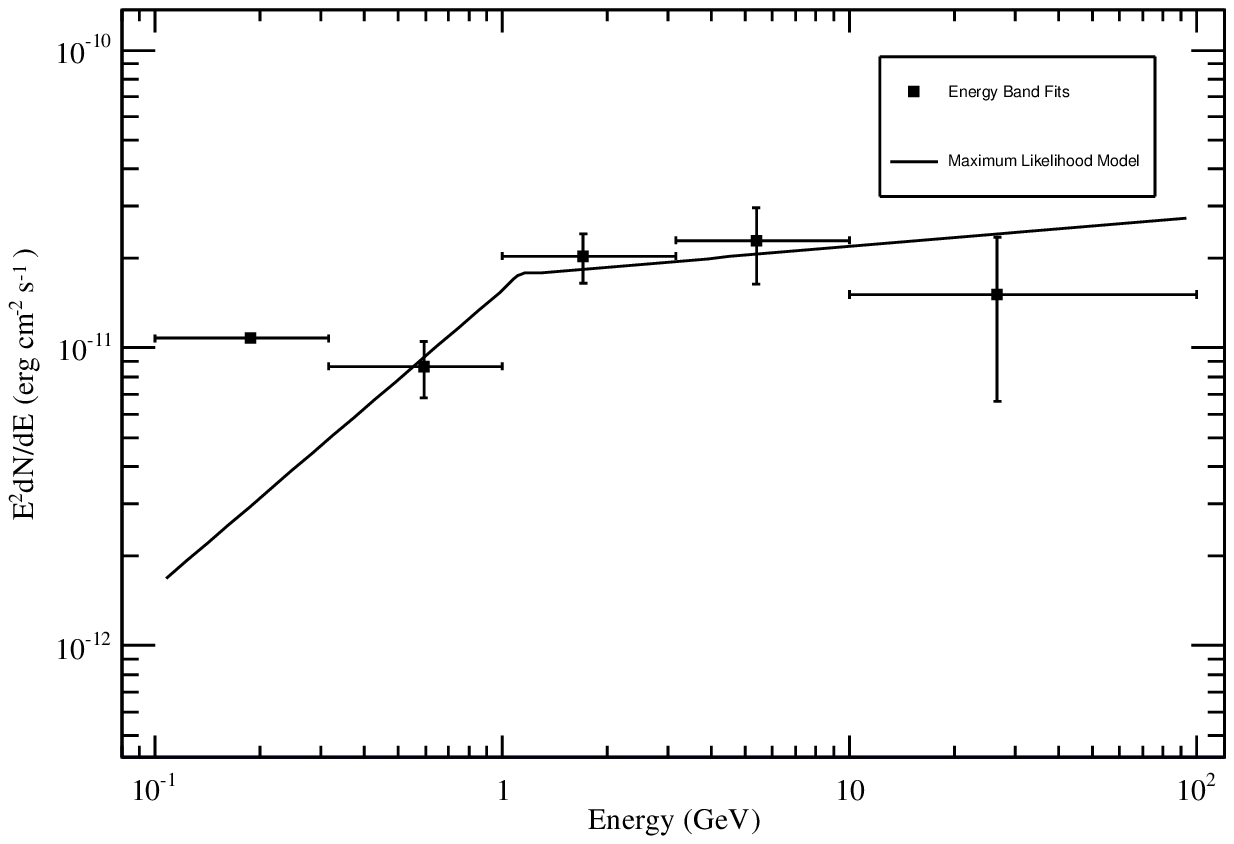}\\

\end{center}
\caption{Spectra for sources from pixels $(a)$ 34, $(b)$ 431, $(c)$ 2663, $(d)$ 9067. \textit{Left:} TS value for each bin of the spectrum.  \textit{Right}: Spectrum of the source in its high state.} \label{fig:spectra1}
\end{figure}

\begin{figure}
\begin{center}

(a)\includegraphics[width=6.0 cm]{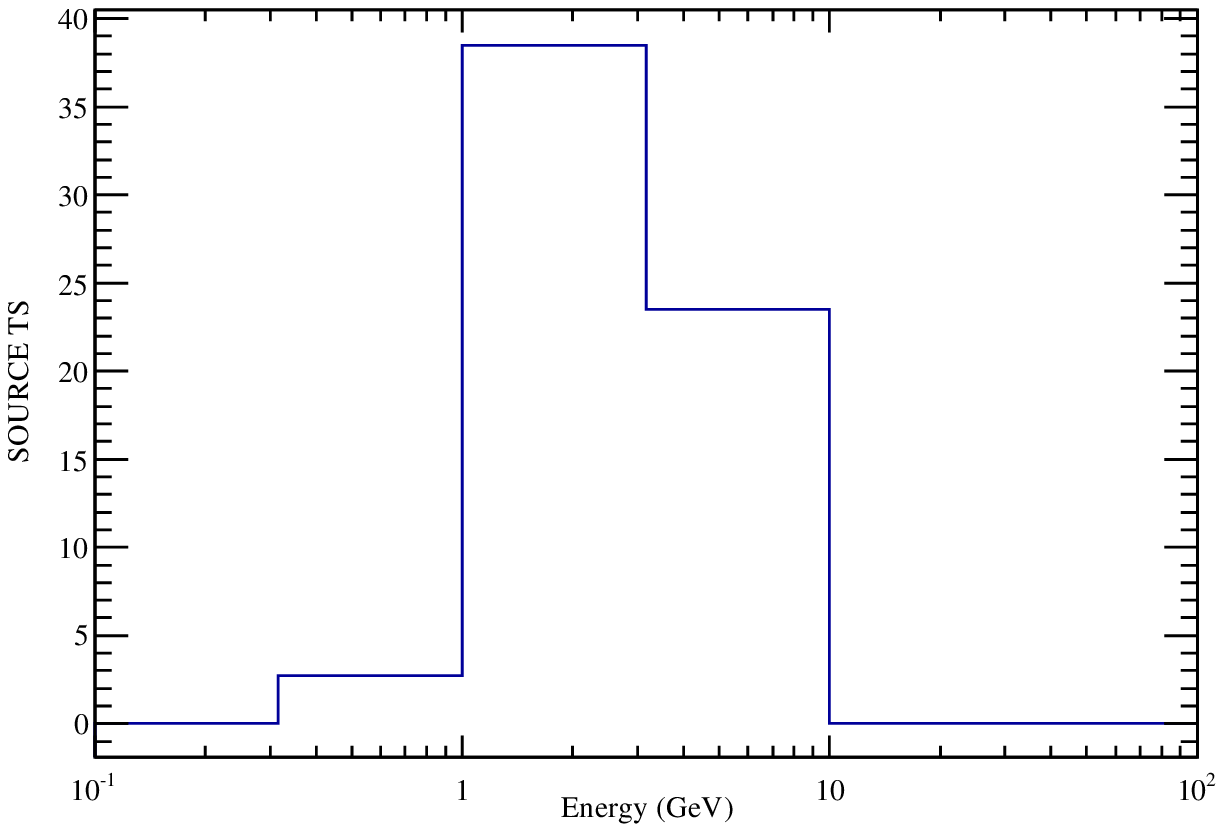} \includegraphics[width=6.0 cm]{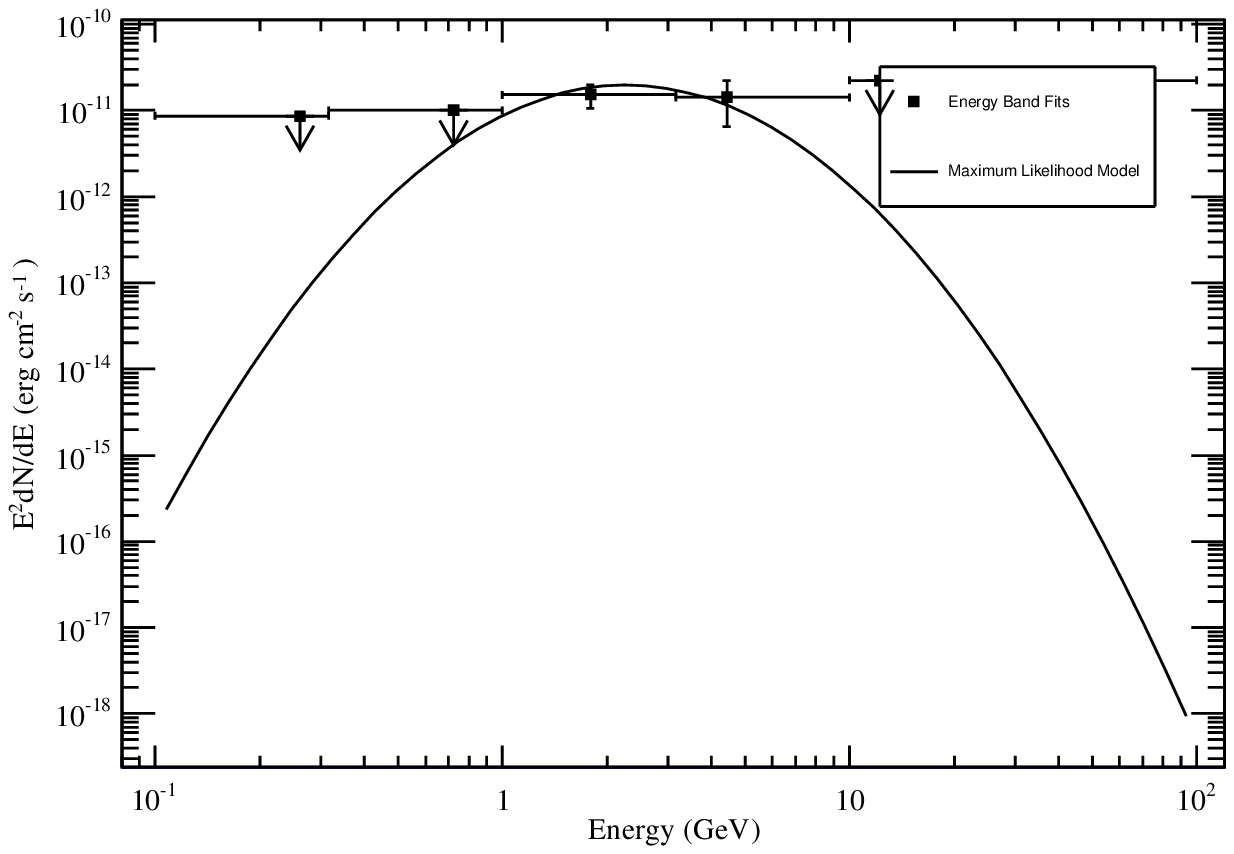}\\
(b)\includegraphics[width=6.0 cm]{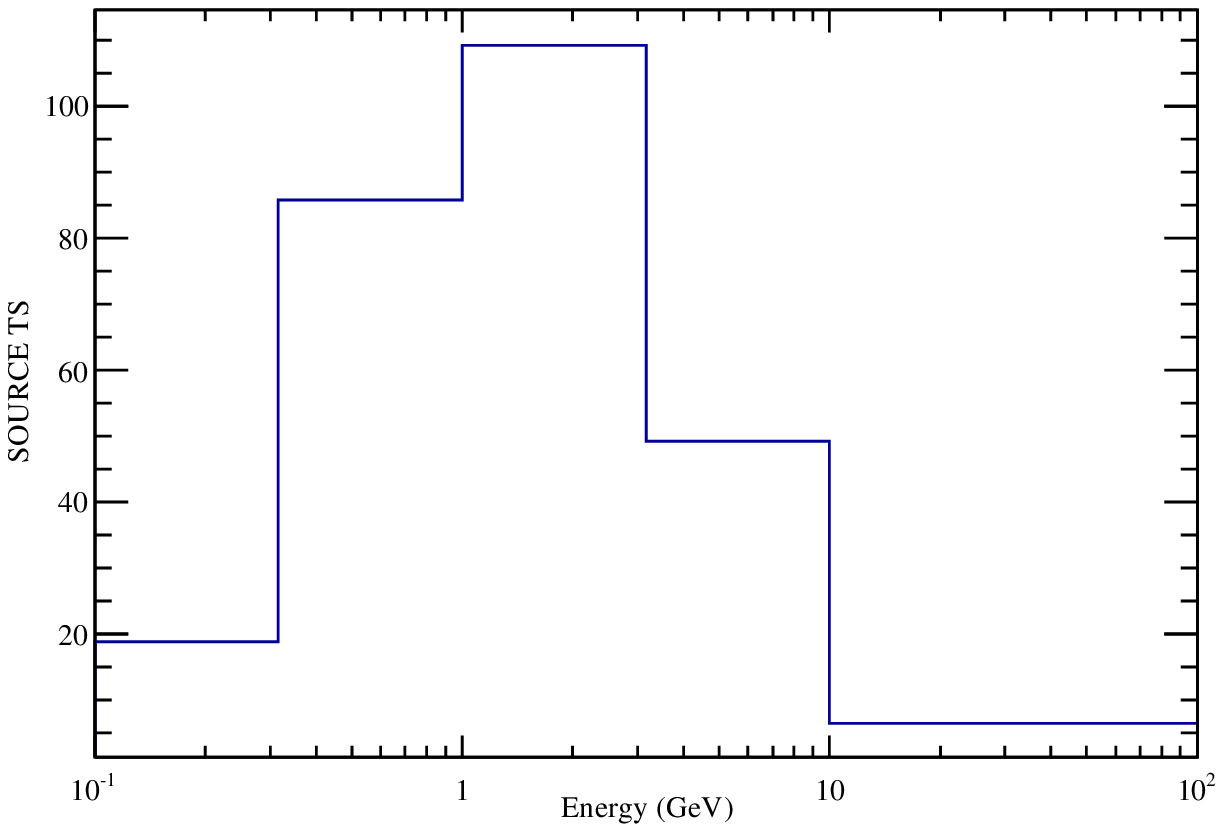} \includegraphics[width=6.0 cm]{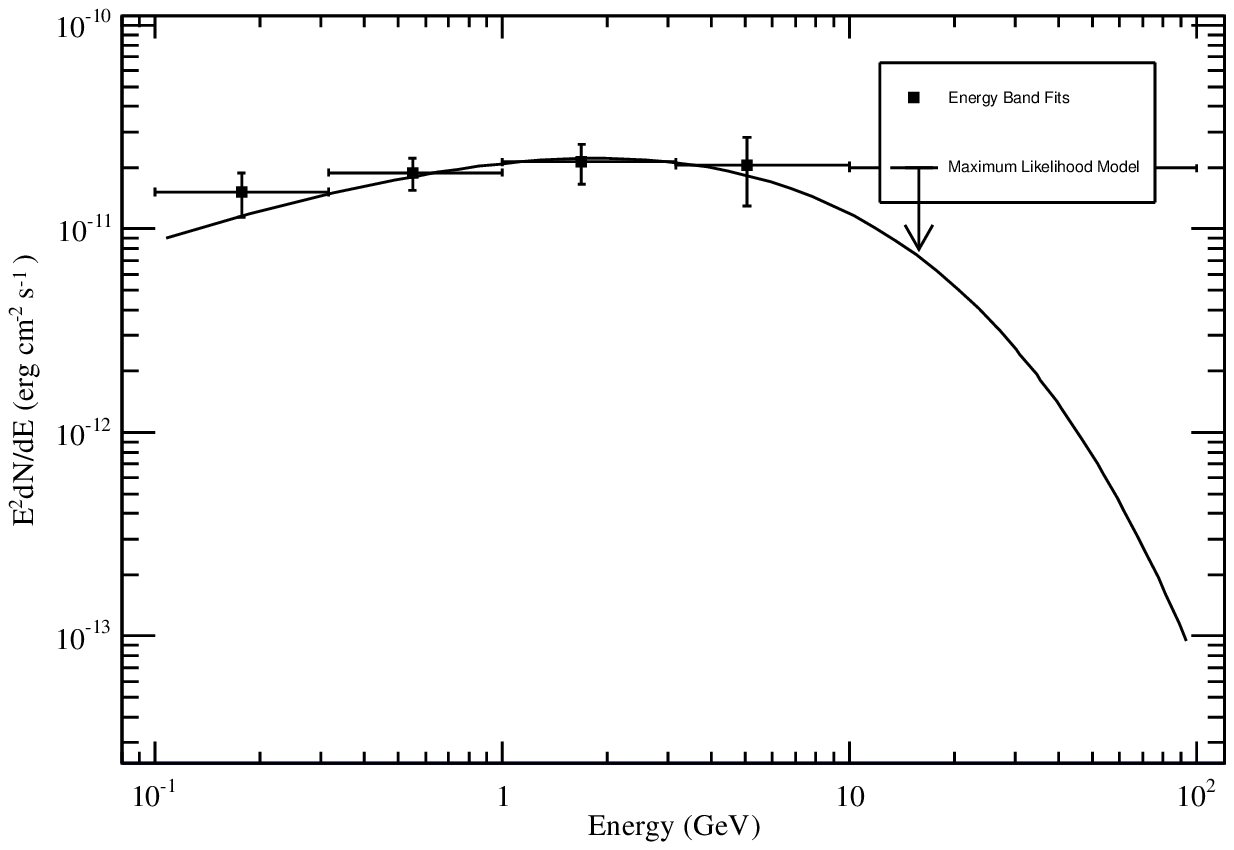}\\
(c)\includegraphics[width=6.0 cm]{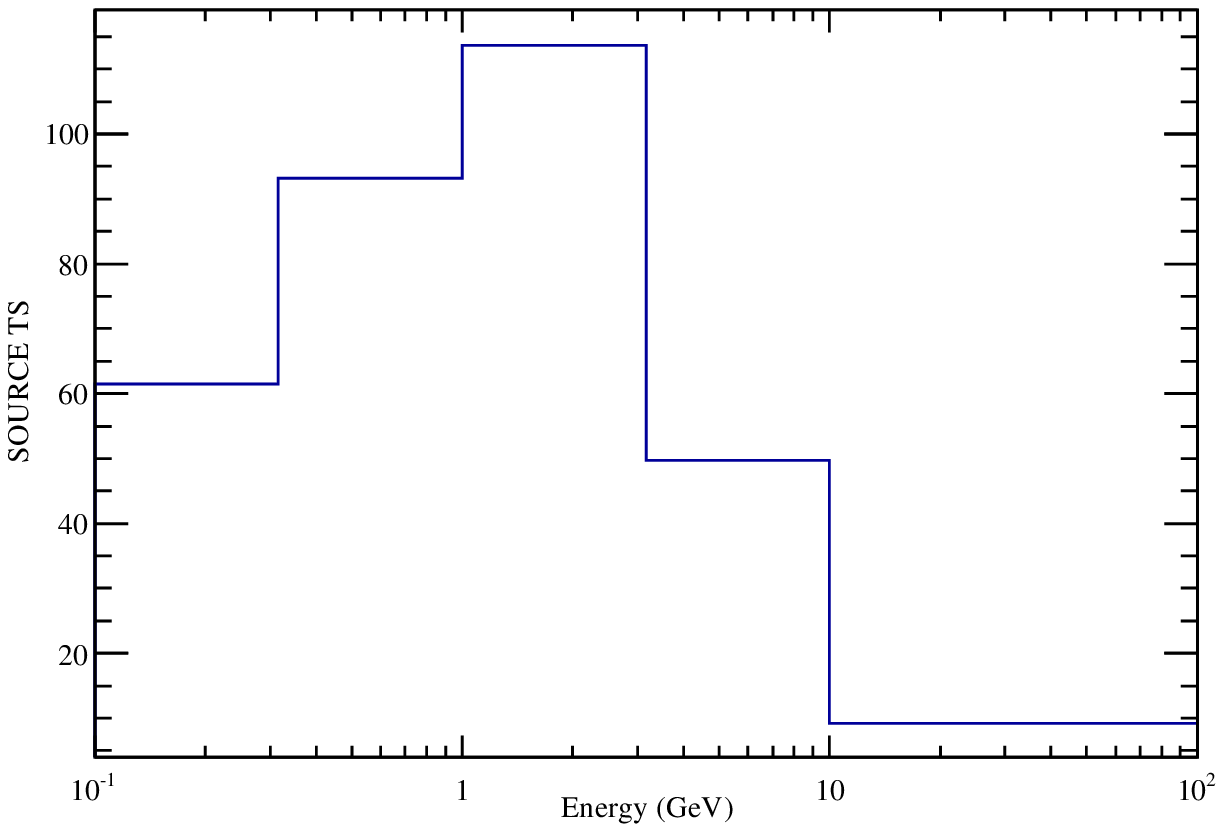} \includegraphics[width=6.0 cm]{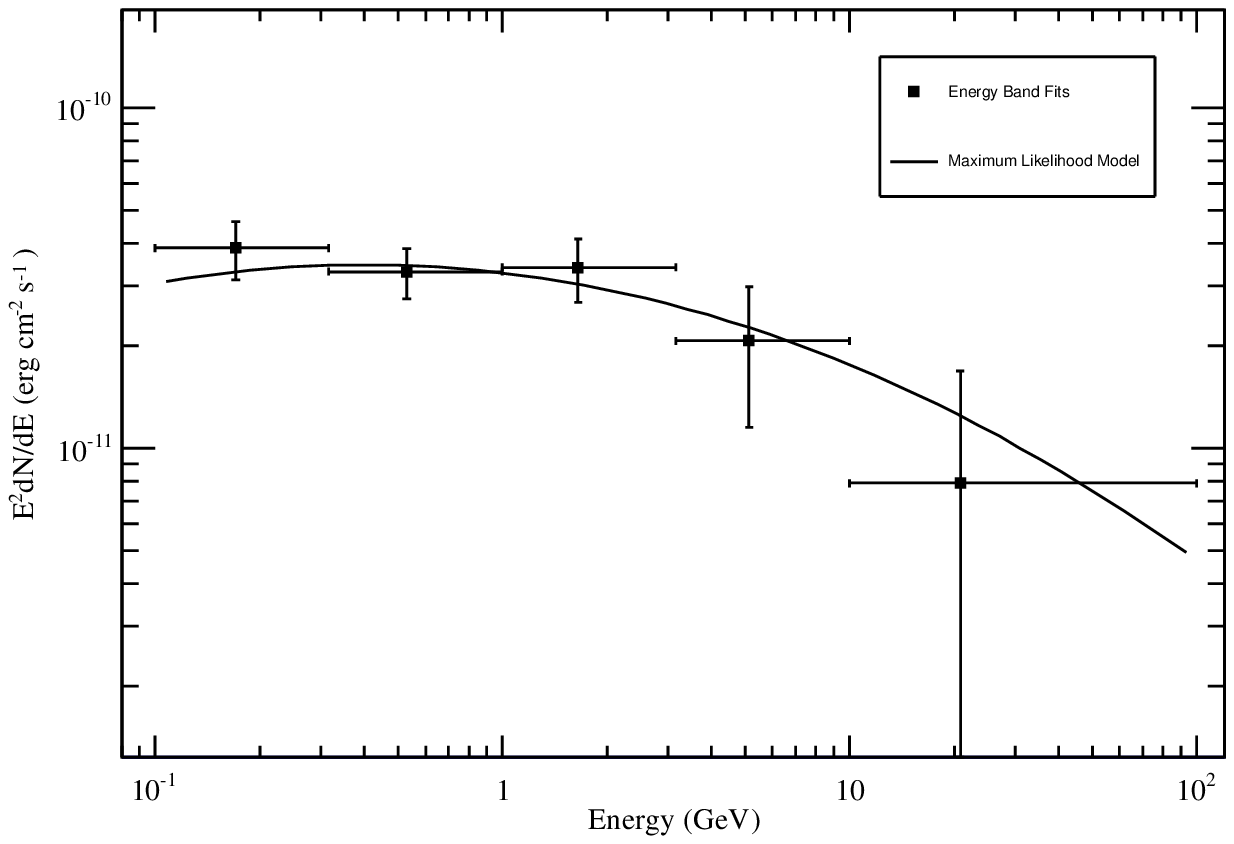}\\
(d)\includegraphics[width=6.0 cm]{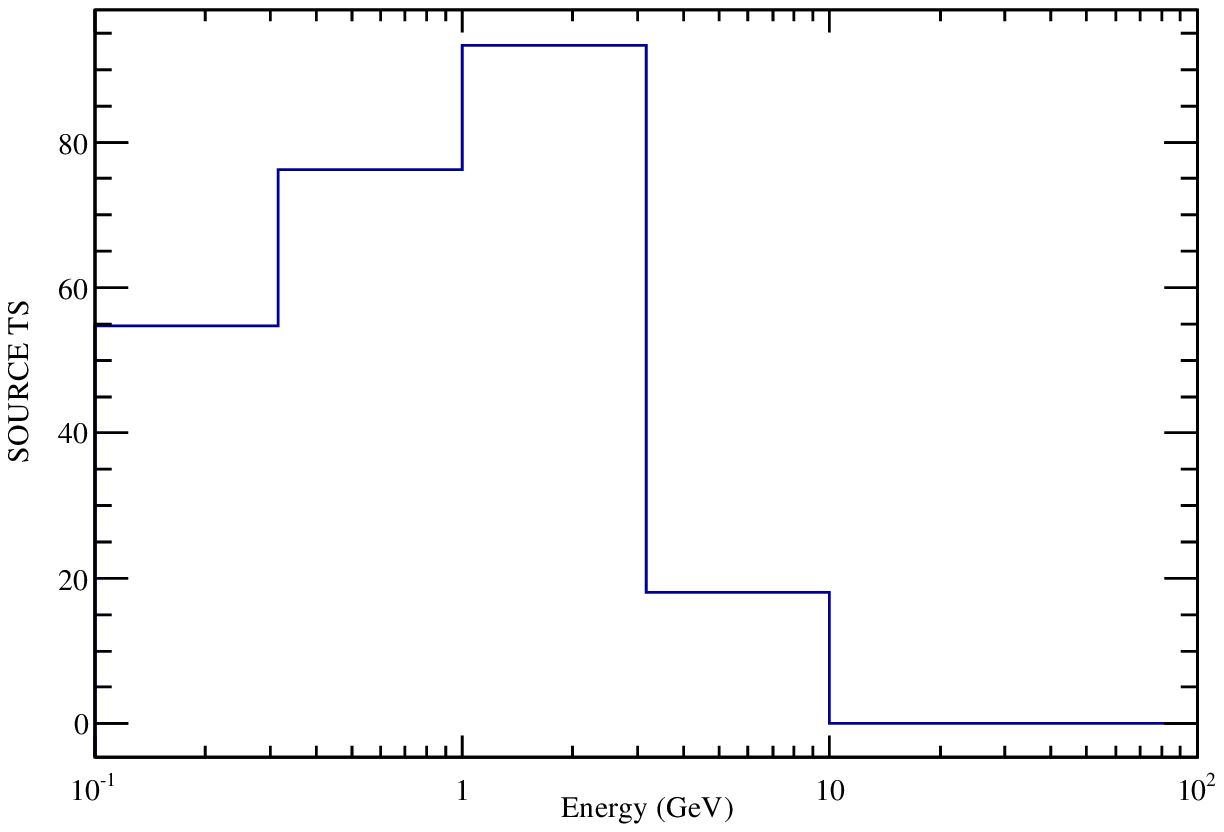} \includegraphics[width=6.0 cm]{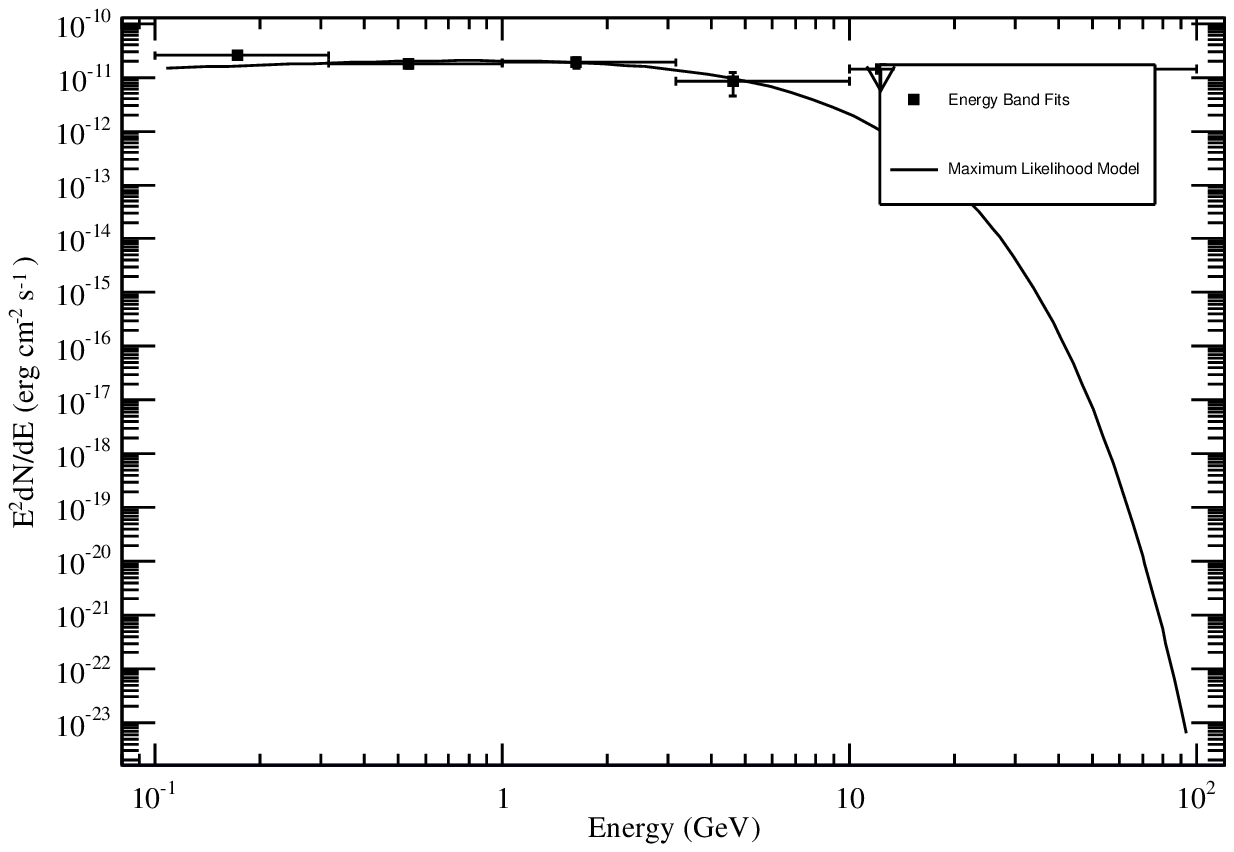}\\

\end{center}
\caption{Spectra for sources from pixels $(a)$ 10871, $(b)$ 11255, $(c)$ 11696, $(d)$ 11901. \textit{Left} TS value for each bin of the spectrum.  \textit{Right}: Spectrum of the source in its high state.} \label{fig:spectra2}
\end{figure}

We have accomplished a detailed analysis of the spectra of these sources. We have
carried out \textit{gtlike} fit in  two energy intervals -- (0.1-100
GeV) and (1.0-100 GeV) adding new sources to the source model files
obtained from the 2FGL catalogue. Also we conduct fits in (0.1-100
GeV) interval using the spectral models other than just a simple
power-law.  The results are presented in the Tables
\ref{table:spectrum},\ref{table:spectra_models}. We plot the spectral
energy distributions (SEDs) using the python-based package likeSED.py\footnote{http://fermi.gsfc.nasa.gov/ssc/data/analysis/user/}(figs. (\ref{fig:spectra1}-\ref{fig:spectra2})). The spectral
shapes demonstrate significant deviation from the simple power law.
For a majority of sources their SEDs peak at around GeV. In all cases
the spectra are rather hard -- that makes them look very similar to
the SEDs of the brightest flares of powerful blazars \citep{Pacciani2014}. Thus the observed variability seems to be caused by some sources with temporal and spectral properties close to highly-variable blazars. 

\section{Conclusions}

We have performed a full sky blind search for regions with variable
flux at energies above 1 GeV in the 308 weeks of the Fermi-LAT
data. The variability of 8 gamma-ray sources was identified for the
first time. Two of the sources are included in the second Fermi LAT source
catalogue (2FGL), one was reported by the EGRET and in the First Fermi
LAT source catalogue (1FGL), while the gamma-ray radiation from the
five other sources is detected for the first time.  The
latter, using the NED and SIMBAD databases, were linked to the sources known in radio and
X-ray bands. All detected variable sources demonstrate recurring flaring activity at
time scales $\sim$ weeks which is typical for blazars. The SEDs of the
sources in their high states are hard and deviate significantly from
the simple power-law. The spectra of the most of the sources peak
around  few GeV. We note finally, that the variability property
was for the first time employed to detect  new gamma-ray sources.

\section*{Acknowedgements}
The work was supported by the RFBR grant 13-02-01293.  M.P. and G.R. acknowledge the fellowships of the Dynasty
foundation. The analysis is based on data and software provided by
the Fermi Science Support Center (FSSC). The numerical part of the
work was done at the cluster of the Theoretical Division of INR
RAS. This research has made use of NASA's Astrophysics Data
System, NASA/IPAC Extragalactic Database (NED) which is operated by
the Jet Propulsion Laboratory, California Institute of Technology,
under contract with the National Aeronautics and Space Administration,
and the SIMBAD database, operated at CDS, Strasbourg, France.
\section*{Update}
When this paper was already finished, the Fermi LAT Third Source Catalog \citep{3FGL} has appeared on the arXiv. We have  checked for our variable sources from the Table \ref{table:undetected} there. 
Both 2FGL sources (pix.34, pix.9067) are in the 3FGL catalogue as well. Sources that were responsible for the variability in the pixels 11255, 11696, 11901 (PKS 0226-559,PKS B0027-024, PMN J0225-2603) are firmly estabisished in the new catalogue. However, 3EG J1424+3734 (pix.431)  is absent from the catalogue. Also our identification in the pix. 2663 (1RXS J161939.9+765515) is 1.3$^{\circ}$ away from the closest 3FGL source, which in turn lacks identification. Our identification of the source  in the pix. 10871 PMN J2320-6447 differs from the 3FGL -- PMN J2321-6438 ($l=317.6567  ^{\circ},b=-49.8059^{\circ}$) which is 26.1$'$ away from our best localization position ($l=317.64^{\circ},b=-49.67^{\circ}$) that has an error 5.5$'$. 
\bibliographystyle{mn2e}

\begin{table}
\begin{center}
\begin{tabular}{|c|c|c|c|c|}
\hline
no&Pixel no. & $l^{\circ}$ & $b^{\circ}$ & error, $'$  \\
1&431&  67.33     & 68.10 & 3.0  \\
2&2663&  110.32  &  34.47& 6.6  \\
3&10871&   317.64    &  -49.67& 5.5  \\
4&11255&   278.42      &  -56.52& 4.0  \\
5&11696&  110.79         &  -64.59 & 3.0	  \\
6&11901&  215.54         &  -69.14 & 3.5	  \\
\hline

\end{tabular}
\end{center}
\caption{Positions  of the  gamma-ray sources not included in the 2FGL that were obtained with the \textit{gtfindsrc} utility. }
\label{table:positions}
\end{table}

\onecolumn

\begin{table}
\begin{center}
\begin{tabular}{|c|c|c|c|c|c|c|c|c|c|}
\hline
no&Pixel no. & \pbox{8 cm}{$F_{100}, ~10^{-8}$\\$\mathrm{ph~cm^{-2}~s^{-1}}$} &\pbox{8 cm} { $\delta F_{100},~10^{-8}$\\$\mathrm{ph~cm^{-2}~s^{-1}}$} & $\Gamma_{100}$& $\delta\Gamma_{100}$&\pbox{8 cm} { $F_{1000},~10^{-8}$\\$\mathrm{ph~cm^{-2}~s^{-1}}$} & \pbox{8 cm} {$\delta F_{1000},~10^{-8}$\\$\mathrm{ph~cm^{-2}~s^{-1}}$} & $\Gamma_{1000}$& $\delta\Gamma_{1000}$ \\
\hline
1&34& $3.7$& $0.04$ & 1.71 &0.004&$0.79$&$0.04$&2.00&0.06 \\

2&431& $3.5$& 0.03 & 1.710 &0.005&0.85&0.05&2.07&0.04 \\

3&2663& 0.68& $0.03$ & 1.72 &0.01&0.14&0.05&2.18&0.34 \\

4&9067& 5.3& 0.07 & 1.68 &0.006&1.2&0.06&1.97&0.03 \\

5&10871& 2.3& 0.07 & 1.64 &0.01&0.75&0.19&2.44&0.30 \\

6&11255& 8.5& 0.2 & 1.99 &0.007&0.95&0.19&2.40&0.29 \\

7&11696& 22& 2.9 & 2.11 &0.098&0.75&0.19&2.44&0.30 \\

8&11901& 12& 0.1 & 2.15 &0.003&0.95&0.16&2.81&0.26 \\
\hline
\end{tabular}
\end{center}
\caption{Spectral properties of the sources in their high states. $F_{100},~F_{1000}$ are photon fluxes at energies higher than 100(1000) MeV, $\delta F_{100},~\delta F_{1000}$ are respective errors on these fluxes, $\Gamma_{100},~\Gamma_{1000}$ are spectral indices in the simple power-law model at these energies, and $\delta\Gamma_{100},~\delta\Gamma_{1000}$ are their errors.}\label{table:spectrum}
\end{table}

\begin{table}
\begin{center}
\begin{tabular}{|c|c|c|c|c|c|}
\hline
no&Pixel no. & PL & BPL & LP & PLEC \\
\hline
1&34& 134.8 & \textbf{137.6} & 136.5 & 136.5 \\

2&431& 159.0 & 162.9 & 165.7 & \textbf{167.5} \\

3&2663& 25.0 & 25.4 & 26.2 & \textbf{26.5} \\

4&9067& 260.1 & \textbf{264.9} &  262.7 & 261.0\\

5&10871& 43.2 & 52.6 & \textbf{58.6} & 56.9 \\

6&11255& 244.6 & 248.4 & 249.1 & \textbf{250.2} \\

7&11696& 297.9 & 298.8 & 298.5 & \textbf{299.1} \\

8&11901& 203.2 & 208.4 & 208.3 & \textbf{210.0} \\
\hline
\end{tabular}
\end{center}
\caption{Goodness-of-fit ($TS$ obtained during $gtlike$ fitting)  with different spectral models: simple power-law (PL), broken power-law (BPL), log-parabola (LP), and power-law with  exponential cut-off(PLEC). The best model for each source is highlighted in bold. }\label{table:spectra_models}
\end{table}

\begin{table}
\begin{center}
\begin{tabular}{|c|c|c|c|c|c|c|c|c|c|}
\hline
no&Pixel no. & $l^{\circ}$ & $b^{\circ}$ & $N_{\mathrm{phot}}$ & $P_{KS}$ &
source identification & distance, $'$ &source type&In 2FGL  \\
\hline

1& 34& 236.25& 84.15& 610&  $1.2\times10^{-24}$& FBQS J122424.1+243623& - & TeV BLL& yes\\
2& 431&  67.33 & 68.10& 419&  $2.2\times10^{-8}$& 3EG J1424+3734& 2 &BLL&no \\
3& 2663& 110.32   & 34.47& 574& $8.5\times10^{11}$& 1RXS J161939.9+765515&10&?&no \\
4& 9067& 120.94&-28.63& 840&  $7.9\times10^{-12}$& GB6 J0043+3426& - &blazar&yes \\
5&10871 &  317.64  & -49.670 &  351 & $1.6\times10^{-7}$   & PMN J2320-6447& 0.2& ?&no\\
6&11255 &  278.42  & -56.52 &  55 &$5.9\times10^{-27}$   & PKS 0226-559& 7 &blazar &no \\
7& 11696& 110.79& -64.59& 326& $1.3\times10^{-9}$&  PKS J0030-0211& 2 &flat-spectrum radio source&no   \\
8&11901 & 215.54   &-69.14&  289&   $3.5\times10^{-6}$ & PMN J0225-2603& 7   &flat-spectrum radio source&no\\
 \hline
\end{tabular}
\end{center}
\caption{List of the sources with previously unreported variability. The coordinates of the sources are taken either from the 2FGL or from the results of the \textit{gtfindsrc} searches. $N_{phot}$ is the number of photons observed in the pixel during 308 weeks,  $P_{KS}$ is the probability that the pixel is non-variable, source identifications were made using the NED and SIMBAD databases.  
  }\label{table:undetected}
\end{table}

\end{document}